# Giant Valley-Zeeman Splitting from Spin-Singlet and Spin-Triplet Interlayer Excitons in WSe$_2$/MoSe$_2$ Heterostructure

Tianmeng Wang[1#], Shengnan Miao[1#], Zhipeng Li[1#], Yuze Meng[1#], Zhengguang Lu[2,3], Zhen Lian[1], Mark Blei[4], Takashi Taniguchi[5], Kenji Watanabe[5], Sefaattin Tongay[4*], Dmitry Smirnov[2], Su-Fei Shi[1,6*]

1. Department of Chemical and Biological Engineering, Rensselaer Polytechnic Institute, Troy, NY 12180, USA
2. National High Magnetic Field Lab, Tallahassee, FL, 32310
3. Department of Physics, Florida State University, Tallahassee, Florida 32306, USA
4. School for Engineering of Matter, Transport and Energy, Arizona State University, Tempe, AZ 85287, USA
5. National Institute for Materials Science, 1-1 Namiki, Tsukuba 305-0044, Japan.
6. Department of Electrical, Computer & Systems Engineering, Rensselaer Polytechnic Institute, Troy, NY 12180, USA

[#] These authors contributed equally to this work
[*] Corresponding authors: shis2@rpi.edu, Sefaattin.Tongay@asu.edu

## ABSTRACT

Transition metal dichalcogenides (TMDCs) heterostructure with a type II alignment hosts unique interlayer excitons with the possibility of spin-triplet and spin-singlet states. However, the associated spectroscopy signatures remain elusive, strongly hindering the understanding of the Moiré potential modulation of the interlayer exciton. In this work, we unambiguously identify the spin-singlet and spin-triplet interlayer excitons in the WSe$_2$/MoSe$_2$ hetero-bilayer with a 60-degree twist angle through the gate- and magnetic field-dependent photoluminescence spectroscopy. Both the singlet and triplet interlayer excitons show giant valley-Zeeman splitting between the K and K' valleys, a result of the large Landé g-factor of the singlet interlayer exciton and triplet interlayer exciton, which are experimentally determined to be ~ 10.7 and ~ 15.2, respectively, in good agreement with theoretical expectation. The PL from the singlet and triplet interlayer excitons show opposite helicities, determined by the atomic registry. Helicity-resolved photoluminescence excitation (PLE) spectroscopy study shows that both singlet and triplet interlayer excitons are highly valley-polarized at the resonant excitation, with the valley polarization of the singlet interlayer exciton approaches unity at ~ 20 K. The highly valley-polarized singlet and triplet interlayer excitons with giant valley-Zeeman splitting inspire future applications in spintronics and valleytronics.

Due to the reduced screening in two-dimension (2D), the enhanced Coulomb interaction not only gives rise to strongly bound exciton in monolayer TMDC[1–6] but also leads to robust interlayer exciton in a TMDC hetero-bilayer of the type II alignment, with the optically excited electron and hole residing in different TMDC layers[7–16]. The spatial separation of the electron and hole results in the long lifetime of the interlayer exciton[17–22]. It was also theoretically predicted that the TMDC hetero-bilayer would host interlayer exciton fine structures: singlet and triplet interlayer excitons[23], a result of the spin-orbit coupling induced splitting of the conduction bands for monolayer TMDCs[24–31]. Interestingly, the triplet interlayer exciton, unlike its counterpart in monolayer TMDC[32,33], is not restricted by the mirror symmetry and could have finite radiation through the in-plane dipole, depending on the atomic registry of the hetero-bilayer[23]. The long-lived singlet and triplet interlayer excitons could also retain valley polarization[34–40], promising for valleytronics applications.

Interestingly, the small twist angle or lattice mismatch between the TMDCs hetero-bilayer would generate a Moiré superlattice potential[41–44], which modulates the interlayer exciton and leads to more fine features. The Moiré modulated interlayer exciton complicates the optical spectra[45,46] and renders it even more challenging to identify the triplet and singlet interlayer excitons. On the other side, identification of the singlet and triplet interlayer excitons would be critical for exploring Moiré potential to engineer interlayer exciton. In this work, we construct a perfectly aligned $WSe_2/MoSe_2$ heterostructure with a 60-degree twist angle. We unambiguously identify the singlet and triplet interlayer excitons through gate-, temperature-, and magnetic field-dependent photoluminescence (PL) spectroscopy. We found that both the singlet and triplet interlayer excitons have a large response to the out-of-plane magnetic field, with the Landé g-factor as large as ~ 10.7 and ~ 15.2, respectively, in excellent agreement with theoretical expectation[47,48]. These g-factors are much larger than that of the bright intralayer exciton in monolayer TMDC (~ 4)[6,49] and lead to a giant valley-Zeeman splitting between the K and K' valleys[50], ~ 11.2 meV and ~ 16.0 meV for the singlet and triplet interlayer excitons, respectively, for an out-of-plane magnetic field of 17.5 T.

We also find that the singlet and triplet interlayer excitons possess opposite valley polarization, corresponding to a $H_h^h$ atomic registry predicted theoretically[23]. The helicity-resolved photoluminescence excitation (PLE) spectroscopy study shows that the PL from each interlayer exciton is highly valley polarized at resonance excitation, which for singlet interlayer exciton can be as high as ~ -100% at 23 K and ~ -85% even for the elevated temperature of 82 K. The high valley polarization of the interlayer excitons originates from the long valley lifetime of the holes[51–53]. The robust valley polarization of the long-lived interlayer excitons thus presents new quasiparticles for valleytronics applications, while the giant valley-Zeeman splitting could be further exploited to break the valley degeneracy.

The monolayer $MoSe_2$ and $WSe_2$ were exfoliated separately, with the crystal orientation determined by the second harmonic generation (SHG). The twist angle between the constructed hetero-bilayer was further determined to be 60 degrees by comparing the

SHG signals from the heterostructure region and monolayer region (see SI). The heterostructure was further encapsulated with few-layer boron nitride (BN) flakes and integrated into the dual gated device using a similar method as described previously[6,49,54]. The schematic of constructed heterostructure device is shown in Fig. 1a, and a typical device image is shown in Fig. 1b. At 77 K, the PL spectra of one device (Fig. 1c) show much quenched PL of intralayer bright exciton of $MoSe_2$ and $WSe_2$, at 1.624 eV and 1.706 eV respectively, while two pronounced interlayer exciton-PL peaks centered at 1.392 eV and 1.416 eV. These two peaks can be tuned by the gate voltage. As shown in Fig. 1d and 1e, when the gate voltage is positive, which corresponds to the electron doping of the bottom layer, the two peaks exhibit a sensitive shift as a function of the top gate voltage due to the Stark shift[54,55]. It is worth noting that, for a device which we place $MoSe_2$ as the bottom layer (Fig. 1d), we observe a blue-shift of the two peaks, opposite to the red-shift observed in the device with the opposite hetero-bilayer stacking order (Fig. 1e), confirming that the observed two PL peaks are associated with the interlayer exciton, whose dipole direction switches as the stacking order switches.

To reveal the nature of the observed two PL peaks associated with the interlayer exciton, we performed a magnetic field dependent PL spectroscopy study. We use a linearly polarized continuous wave (CW) laser centered at 1.959 eV as the excitation, and the obtained PL spectra as a function of the out-of-plane magnetic field are shown as a color plot in Fig. 2a. It is evident that, in the presence of the magnetic field, each of the PL peaks is split into two, with the splitting as large as ~16.0 meV for the triplet interlayer exciton ($IX_T$) and ~11.2 meV for the singlet interlayer exciton ($IX_s$) under the magnetic field of 17 T. This splitting is due to the valley-Zeeman shift since the linearly polarized light excites both K and K' valleys, which undergoes an opposite shift in the presence of the out-of-plane magnetic field[26–29]. This splitting can be expressed as $\Delta E = g\mu_B B$, in which g is the Landé g-factor, $\mu_B$ is the Bohr magneton, and B is the magnetic field strength. As shown in Fig. 2b, the experimentally measured energy splitting for different magnetic field strengths can be fitted with a linear function, which gives the g factor ~ 15.2 for $IX_T$ and ~ 10.7 for $IX_s$. These g factors are much larger than that of the bright exciton or trions in monolayer TMDC (~ 4)[6,49,56,57], attributing to the giant valley-Zeeman splitting of the interlayer excitons[58].

The measured g-factor can also be used to determine the nature of the interlayer exciton peaks. As shown by recent studies, the spin-orbit coupling in monolayer TMDC not only leads to a large splitting of the valence bands (300-500 meV) but also gives rise to a splitting in the conduction bands, which is much smaller in scale (~ 20 meV)[59,60]. As a result, for the $WSe_2/MoSe_2$ heterostructure with a 60-degree twist angle, the K valley of $WSe_2$ is aligned with the K' valley of $MoSe_2$, as schematically shown in Fig. 2c, d. It has been shown previously that the conduction band splitting gives rise to the spin triplet exciton, aka spin-dark exciton, in monolayer TMDCs[32,33,58]. Here, the presence of the conduction band splitting renders it possible to have two configurations of the interlayer exciton: spin triplet ($IX_T$) and spin singlet ($IX_S$) interlayer exciton, as shown in Fig. 2d. At a finite temperature, the thermal fluctuation ensures a certain population of exciton in the higher energy IX$_S$, and both peaks are clearly visible in the PL spectra (Fig. 1d-e and Fig. 2a). In a non-interacting picture, the g factor for either of the interlayer exciton, IX$_T$ or IX$_S$, can be calculated theoretically counting the overall contribution of the spin, orbital and valley components, and it is expected to be 16 for IX$_T$ and 12 for IX$_S$ (see SI). The experimentally extracted values are in excellent agreement with the theoretical expectations, confirming the assignment of the singlet and triplet interlayer excitons. It is worth noting that, in monolayer WSe$_2$ or WS$_2$, the triplet exciton is a spin-forbidden dark exciton and can only radiate through an out-of-plane dipole[6,31,49,61–65]. In the WSe$_2$/MoSe$_2$ heterostructure that we investigate here, due to the lift of the out-of-plane mirror symmetry, the triplet interlayer exciton is not necessarily dark and could have significant in-plane dipole radiation. The recent theory has shown that, depending on the exact atomic registry of the hetero-bilayers, there could be significant in-plane dipole radiation with valley information retained[23].

To investigate the valley polarization of the interlayer excitons, we perform helicity resolved PLE spectroscopy study. We first excite the heterostructure with circularly polarized light ($\sigma^+$) with different the excitation photon energies, and we detect the PL with the same ($\sigma^+$) or opposite helicity ($\sigma^-$). The obtained data near the interlayer exciton energy (from 1.355 eV to 1.470 eV) at 77 K is shown in Fig. 3a. It is evident that the PL intensity is at its local maximum when the excitation is in resonance with the A or B exciton of either MoSe$_2$ ($X_A^{Mo}$, $X_B^{Mo}$) or WSe$_2$ ($X_A^{W}$, $X_B^{W}$). More interestingly, at resonant excitation, the relative intensity of the IX$_S$ and IX$_T$ in different detection schemes are significantly different, as shown in Fig. 2d. For the excitation photon energy (1.722 eV) in resonance with the A exciton of the WSe$_2$, the higher PL intensity peak of IX$_T$ in the $\sigma^+\sigma^+$ detection scheme switches to the lower intensity in the $\sigma^+\sigma^-$ configuration. Similarly to the definition of the intralayer bright exciton in monolayer TMDC, we define valley polarization as $P = \frac{I(\sigma^+)-I(\sigma^-)}{I(\sigma^+)+I(\sigma^-)}$, in which $I(\sigma^+)$ is the integrated PL intensity of the same helicity ($\sigma^+$) and $I(\sigma^-)$ is the integrated PL intensity of the opposite helicity ($\sigma^-$). The switch of the relative PL intensity is due to the negative valley polarization of the IX$_S$ and positive valley polarization of the IX$_T$ (Fig. 3b). The particular helicity of PL emission agrees with the $H_h^h$ atomic registry (see SI Section S7)[23], in which the singlet and triplet interlayer exciton both radiate through an in-plane dipole but with opposite helicity. As shown in Fig.3b, the valley polarization of the two interlayer excitons is sensitively dependent on the excitation photon energy. At the excitation photon energy of 1.722 eV, the valley polarization of the singlet interlayer exciton (IX$_S$) is as high as ~ 32.9% and the triplet interlayer exciton (IX$_T$) could be as high as ~ -58.6% at 77 K (Fig. 3d). When the excitation photon energy is off-resonance (2.138 eV in Fig. 3c), the valley polarization for both interlayer excitons vanishes.

The valley polarization also sensitively depends on the temperature. The temperature dependent PL spectra for the $\sigma^+\sigma^+$ and $\sigma^+\sigma^-$ configurations are shown in Fig. 4a and Fig. 4b, respectively. Fig. 4a shows that, as we detect the PL of the same helicity of the excitation light, the higher energy singlet interlayer exciton PL decreases significantly as the temperature is decreased, suggesting that the singlet interlayer exciton is in thermal equilibrium with the triplet interlayer exciton and is thermally populated. The thermal equilibrium picture is also confirmed through lifetime measurements, which shows similar lifetime of the singlet and triplet interlayer exciton (see SI). In contrast, in the $\sigma^+\sigma^-$ scheme (Fig. 4b), as we detect the PL in the opposite helicity channel, the singlet interlayer exciton PL becomes pronounced at the lower temperature, which is due to the increased negative valley polarization and is shown explicitly in Fig. 4c. Fig. 4c plotted the valley polarization (P) of the triplet and singlet interlayer excitons as a function of the temperature at the resonant excitation of 1.722 eV (in resonance with the A exciton of $WSe_2$). As the temperature is decreased to 23 K, the valley polarization of the singlet interlayer exciton is approaching 100%. This high valley polarization arises from the long valley lifetime of the hole[51–53]. The optically excited electron-hole pairs can quickly dissociate and separate into two layers[66–69], but the intervalley scattering of the hole is strongly inhibited[34,50,51], especially for resonance excitation[35–40]. This sensitive temperature dependence of the valley polarization for both interlayer exciton can be easily illustrated in Fig. 4d, which shows PL spectra of the $\sigma^+\sigma^+$ and $\sigma^+\sigma^-$ configurations for temperatures of 42 K and 82 K.

Interestingly, the valley polarization of the triplet interlayer exciton seems to saturate at the temperature around 102 K, with a saturated value of ~ 40%, less than that of the singlet interlayer exciton. We suspect that it is due to the lower energy nature of the triplet interlayer exciton, and the PL is more likely to be affected by defect PL at similar energy. This is supported by our observation in Fig. 4a, b, which shows that the PL near 1.395 eV is significantly broadened as the temperature is decreased.

In summary, we report the unambiguous identification of the triplet and singlet interlayer excitons through gate-, magnetic field-, and temperature-dependent PL spectroscopy study. The helicity resolved PLE spectroscopy study shows that the valley polarization of the interlayer exciton can approach unity at low temperature. These highly valley polarized interlayer excitons, with giant valley-Zeeman splitting, inspire future exploration of applications in valleytronics and spintronics. Utilization of the proximity field of a 2D magnetic material[70–75] could take advantages of the large valley-Zeeman splitting in a device configuration, without the necessity of involving an external magnetic field.

## Supporting Information

Details about sample preparation, optical characterization and data analysis can be found in the supporting information (SI).

## Notes

The authors declare no competing financial interest.

**ACKNOWLEDGMENTS**

We thank Dr. Chenhao Jin for helpful discussions. T. Wang and S.-F. Shi acknowledge support from ACS PRF through grant 59957-DNI10. S. Miao, Z. Li and S.-F. Shi acknowledge support from AFOSR through Grant FA9550-18-1-0312. Z. Lian and S.-F. Shi acknowledge support from NYSTAR through Focus Center-NY–RPI Contract C150117. S. Tongay acknowledges support from NSF DMR-1552220, DMR-1838443, and CMMI-1933214. The device fabrication was supported by Micro and Nanofabrication Clean Room (MNCR) at Rensselaer Polytechnic Institute (RPI). K.W. and T.T. acknowledge support from the Elemental Strategy Initiative conducted by the MEXT, Japan and the CREST (JPMJCR15F3), JST. Z. Lu and D.S. acknowledge support from the US Department of Energy (DE-FG02-07ER46451) for magneto-photoluminescence measurements performed at the National High Magnetic Field Laboratory (NHMFL), which is supported by National Science Foundation through NSF/DMR-1644779 and the State of Florida. S.-F. Shi also acknowledges the support from a KIP grant from RPI and a VSP grant from NHMFL.

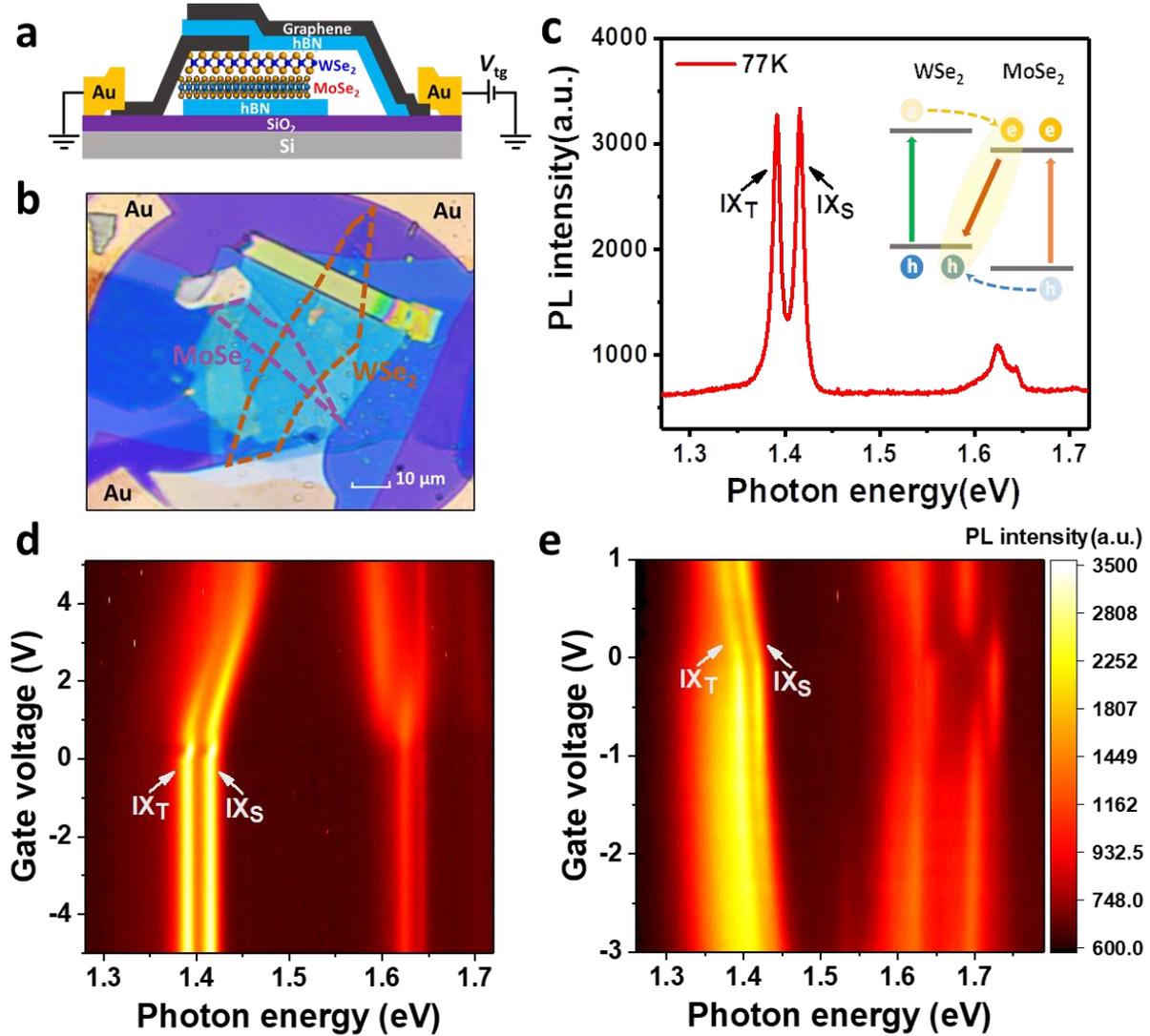

**Figure 1. Interlayer excitons in WSe$_2$/MoSe$_2$ hetero-bilayer.** (a) Schematic of the BN-encapsulated WSe$_2$/MoSe$_2$ heterostructure. One piece of few-layer graphene is used as the contact electrode and another piece is used as the transparent top-gate electrode. (b) Microscope image of the device. Scale bar: 10 μm. (c) PL spectra of the heterostructure, which exhibits two interlayer exciton peaks at 77 K. The CW laser centered at 1.959 eV was used as the excitation source. Inset: schematic of the type II band alignment of WSe$_2$/MoSe$_2$ hetero-bilayer. (d) Color plot of the PL spectra at 77 K of the WSe$_2$ (on top)/MoSe$_2$ heterostructure as a function of gate voltage, which is the same stacking sequence as the scheme in (a). (e) Color plot of the PL spectra at 77 K of the MoSe$_2$(on

top)/WSe$_2$ heterostructure as a function of gate voltage, which is the opposite stacking sequence as the scheme in (a).

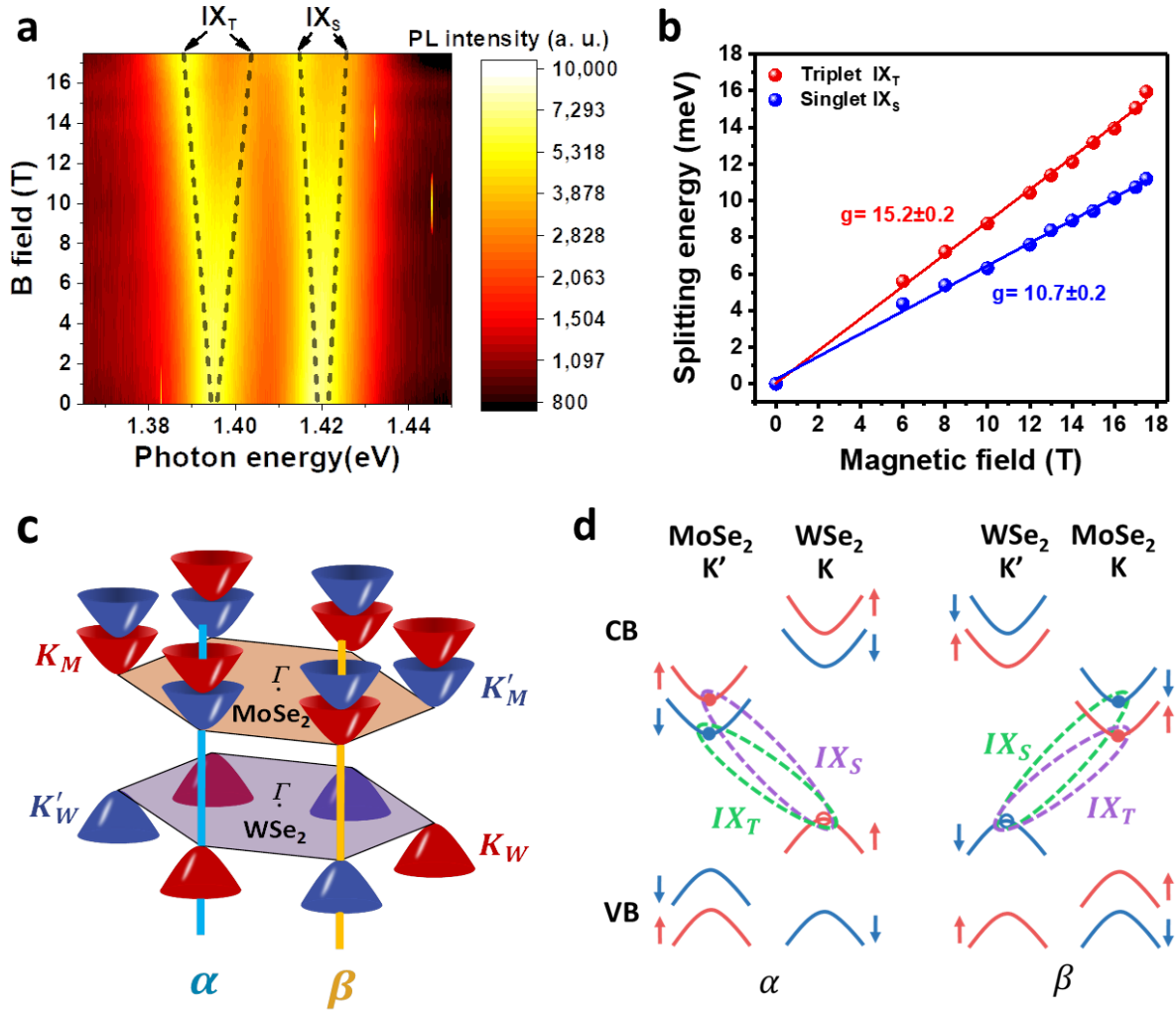

**Figure 2. Magneto-PL spectra of interlayer excitons.** (a) Color plot of the PL spectra of interlayer exciton as a function of the out-of-plane magnetic field at 77 K. The CW laser centered at 1.959 eV with a power of 250 μW was used as the excitation source. (b) PL peak energy splitting for both interlayer exciton states. The valley-Zeeman splitting of each interlayer exciton is utilized to extract the corresponding g-factor through a linear fitting. (c) Illustration of the band structure at the corners (K and K' valleys) of the hexagonal Brillouin zone of a MoSe$_2$/WSe$_2$ heterostructure, with the twist angle of 60-degree. The K and K' valleys at the conduction band minimum (in MoSe$_2$) and valence band maximum (in WSe$_2$) are aligned in momentum space. Here α and β are the heterostructure valleys, and red color stands for spin up, blue color stands for spin down.

(d) Configurations of the interlayer exciton at α and β valleys. Solid dots represent the electrons and the empty ones represent the holes. The dashed lines indicate the formation of triplet interlayer exciton (IX$_T$) and the singlet interlayer exciton (IX$_S$), where green (purple) color represents $\sigma^+(\sigma^-)$ helicity PL observed experimentally.

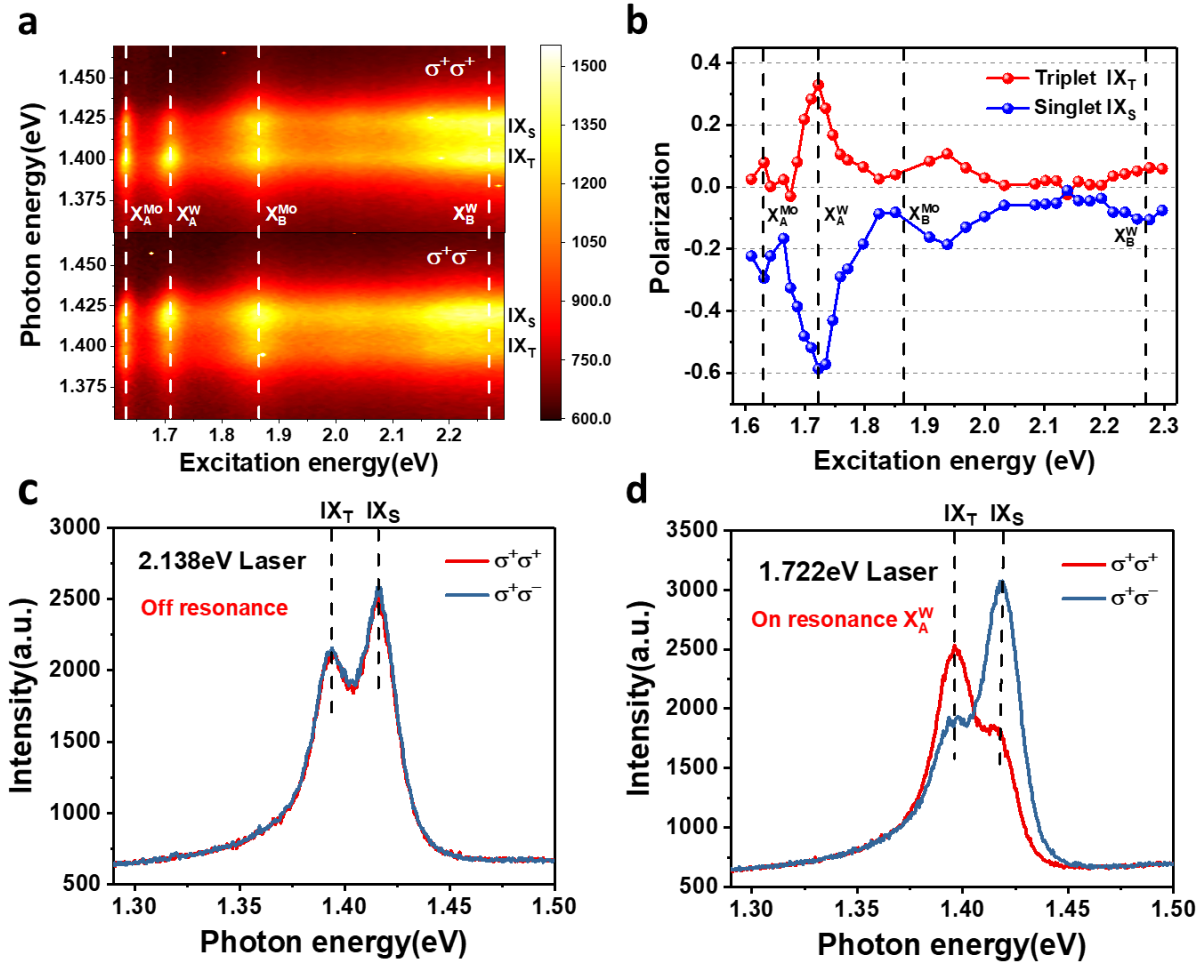

**Figure 3. PLE spectra of interlayer excitons at 77 K.** (a) Color plot of the PL spectra of interlayer exciton as a function of excitation photon energy. $X_B^W$, $X_B^{Mo}$, $X_A^W$, $X_A^{Mo}$ correspond to the photon energy of WSe$_2$ B exciton, MoSe$_2$ B exciton, WSe$_2$ A exciton and MoSe$_2$ A exciton modes, respectively. The PL measurement uses circularly polarized light ($\sigma^+$) for excitation and detects PL with the same ($\sigma^+$) or opposite ($\sigma^-$) helicity. (b) Valley polarization of two interlayer exciton states as a function of excitation energy. (c) PL spectra of interlayer exciton excited off-resonance, with the excitation photon energy centered at 2.138 eV. (d) PL spectra of interlayer exciton excited resonantly at WSe$_2$ A exciton energy (1.722 eV).

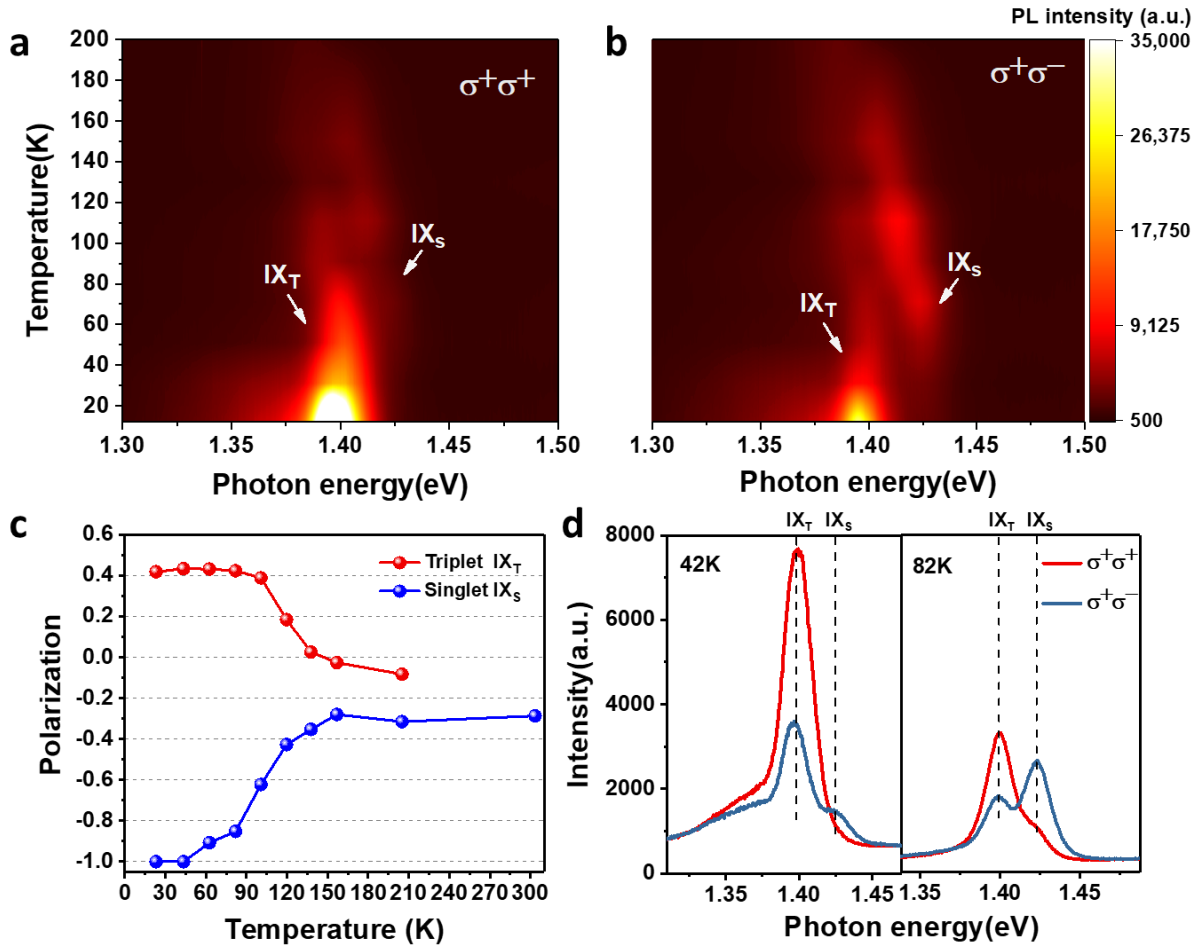

**Figure 4. Temperature-dependent valley polarization of interlayer excitons.** (a) and (b) are color plots of the PL spectra of interlayer exciton as a function of temperature in $\sigma^+\sigma^+$ and $\sigma^+\sigma^-$ configuration, respectively. (c) Valley polarization of two interlayer exciton states as a function of temperature. (d) Representative PL spectra of interlayer exciton at 42 K and 82 K.

TOC

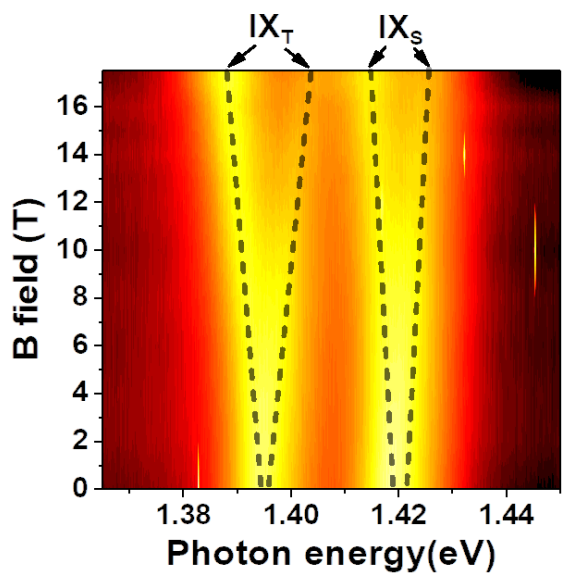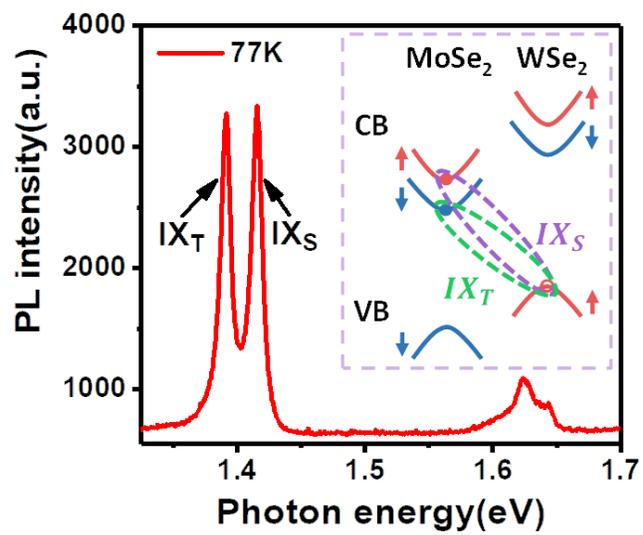